\documentclass[12pt,preprint]{aastex}

\usepackage{amssymb,latexsym,amsmath,graphics}

\slugcomment{}

\shorttitle{MAGNETIC DISKS AROUND NEUTRON STARS}

\shortauthors{ERKUT \& ALPAR}

\begin{document}

\title{ON THE ROTATIONAL DYNAMICS OF MAGNETICALLY THREADED DISKS AROUND NEUTRON STARS}

\author{M. HAKAN ERKUT\altaffilmark{1}}
\affil{Bo\u{g}azi\c{c}i University, 34342, Bebek, \.{I}stanbul,
Turkey; Sabanc\i\ University, 34956, Orhanl\i\--Tuzla,
\.{I}stanbul, Turkey}


\and

\author{M. AL\.{I} ALPAR\altaffilmark{2}}
\affil{Sabanc\i\ University, 34956, Orhanl\i\--Tuzla, \.{I}stanbul,
Turkey}


\altaffiltext{1}{hakane@sabanciuniv.edu}
\altaffiltext{2}{alpar@sabanciuniv.edu}

\begin{abstract}
We investigate the rotational dynamics of disk accretion around a
strongly magnetized neutron star with an aligned dipole field. The
magnetospheric field is assumed to thread the disk plasma both
inside and outside the corotation radius. As a result of disk-star interaction,
the magnetic torque on the disk affects the structure of accretion flow to yield
the observed spin-up or spin-down rates for a source of given fastness, magnetic field strength,
and mass accretion rate. Within the model we obtain a prescription for the dynamical viscosity
of such magnetically modified solutions for a Keplerian disk. We then use this prescription
to find a model solution for the rotation rate profile throughout the entire disk including the
non-Keplerian inner disk. We find that the non-Keplerian angular velocity transition region is
not necessarily narrow for a source of given spin state. The boundary layer approximation,
as in the standard magnetically threaded disk model, holds only in the case of dynamical
viscosity decreasing all the way to the innermost edge of the disk. These results are applied
to several observed disk-fed X-ray pulsars that have exhibited quasi-periodic oscillations (QPOs).
The QPO frequencies provide a constraint on the fastness parameter and enable one to determine uniquely the
width of the angular velocity transition zone for each source within model assumptions.
We discuss the implications of these results on the value of the critical fastness parameter
for a magnetized star in spin equilibrium. Applications of our model are also made with
relevant parameters from recent numerical simulations of quasi-stationary disk-magnetized star interactions.
\end{abstract}

\keywords{accretion, accretion disks --- magnetic fields ---
stars: magnetic fields --- stars: neutron --- X-rays: stars}

\section{INTRODUCTION}

The evidence for the interaction between a rotating magnetized star and a
surrounding accretion disk emerges in a wide variety of astrophysical
systems such as X-ray binary pulsars \citep{JR84}, magnetic white dwarfs in
cataclysmic variables \citep{Warner95}, and T Tauri stars \citep{BB89}. The
detailed modelling of this interaction would contribute to our understanding
of many problems like the time rate of change in the pulse periods of the
disk fed neutron stars and the quasi-periodic oscillations (QPOs) observed
in some sources \citep{AS85,AK90}.

In the early investigations \citep{Sch78,GL78,Aly80},
the degree of diamagnetism presumed for the disk plasma has led to very
different magnetic field configurations. If the disk is fully diamagnetic,
then it excludes the stellar field completely and the accretion flow from
the disk midplane to the surface of the star is assumed to start within a
very narrow region near the inner disk radius \citep{Ich78}. If the accreting
plasma has a non-zero resistivity, then the stellar field may penetrate the
disk and thread it in a broad zone via the Kelvin-Helmholtz instability,
magnetic field reconnection with small scale fields in the disk and
turbulent diffusion \citep{GL79a,Wang87}. One
of the most important predictions of the magnetically threaded disk (MTD)
model is that the star can spin down while matter accretes on it.

In the MTD model of GL79, the star-disk interaction region consists of two
distinct parts. In a broad outer transition zone, where the angular velocity
is Keplerian, the effective viscous stress is dominant compared to magnetic
stress associated with the twisted field lines. The disk matter is brought
into corotation with the neutron star only in a narrow inner transition zone
or boundary layer within which magnetic stress dominates over viscous
stress. The disk resistivity is very high so that the magnetic diffusivity
exceeds the turbulent viscosity by several orders of magnitude. The coronal
plasma outside the disk is force-free and the disk is threaded over a large
range of radii by magnetic field lines that close through the neutron star.

The study of field line twisting in a force-free magnetosphere has revealed
that a closed field line tends to inflate and evolve into an open one
breaking the disk-star link \citep{LB94,LRBK95}.
The strong outflows associated with the opening of
magnetic field lines were reported in several two dimensional
magnetohydrodynamic (MHD) simulations \citep{HSM96,MS97,GWB97,GBW99}.
Although these numerical studies indicated that the accretion flow
around a strongly magnetized object can be nonstationary and episodic in
character, the recent MHD simulations by \citet{RUKL02}
showed that quiescent accretion regimes are also possible
provided a matter dominated differentially rotating corona instead of a
magnetically dominated force-free one is used as an equilibrium initial
condition to examine quasi-stationary situations. The main reason for obtaining
different results in the above mentioned simulations is that in the case of a corona
corotating with the central star (or non-rotating corona), the initial discontinuity
in the angular velocity between the Keplerian disk and the magnetosphere leads
to huge field twisting on the corona-disk boundary. This gives rise to the
generation of very large toroidal field and to strong magnetic braking of the
disk. A differentially rotating matter dominated corona is an appropriate
initial condition to avoid this discontinuity \citep{RUKCL98}. It reduces
the initial magnetic braking of the disk and allows one to investigate
the quiescent phase of disk-star interaction in long-term numerical simulations (RUKL02).

The physical mechanism responsible for the disruption of the inner region of
an axisymmetric disk around a magnetic rotator was investigated by a number
of authors (Brandenburg \& Campbell 1998; Campbell \& Heptinstall 1998, hereafter CH98).
Integrating the relevant MHD equations radially from the weakly magnetic outer disk regions
inwards, they found a vertically disrupted and viscously unstable disk
solution inside a critical radius. They also argued that the viscous
instability associated with the elevated temperature in the inner disk
regions is due to magnetically enhanced viscous dissipation, which causes
the vertical equilibrium to break down when the radiation pressure becomes
significant. We point out in the present paper (see \S \S\ \ref{sec3}
and \ref{sec4}), that the viscous instability proposed by these authors as
the mechanism responsible for the truncation of the inner disk is a
consequence of choosing Keplerian rotation for the disk plasma throughout
the accretion flow, artificial in the sense that the angular motion could
easily be modified by the huge viscous stress estimated for the inner parts
of the disk.

For a disk around a non-magnetized star, Glatzel (1992) solved this
problem without making the assumption that the rotation rate is Keplerian
except in a very narrow boundary layer. In this paper we will present the
analogous solution for the magnetic case, allowing for non-Keplerian
rotation in the inner disk. Numerical simulations by RUKL02 also have shown
non-Keplerian rotation in the magnetically braked inner regions of the disk. We consider
an axisymmetric thin disk threaded by the magnetic dipole field of an aligned rotator
and explore the effect of the rotating magnetosphere on the angular velocity profile
of the disk plasma. In \S\ \ref{sec2} we introduce the relevant MHD
equations to derive a dimensionless function for the vertically integrated
dynamical viscosity. In \S\ \ref{sec3} we discuss the behavior of this
function in connection with the net torque acting on the central star. In 
\S\ \ref{sec4} we employ the dynamical viscosity and solve the equation of
angular momentum conservation for the rotation profile of a disk around a
neutron star in spin equilibrium without making boundary layer
approximation. We apply our model to several X-ray binary pulsars that have
exhibited QPOs. Using the observed periods, X-ray luminosities, QPO
frequencies, and spin-up or spin-down rates of these sources, we obtain
constraints on the model parameters within the beat frequency model and
estimate the appropriate disk rotation curve for each source. We find that
the inner disk rotation rate adapts to Keplerian rotation in the outer disk
through a transition zone whose width and rotation rate profile are such as
to accomodate the mass accretion rate, rotation period, and magnetic moment
of the central object and to yield the observed spin-up or spin-down rates.
In \S\ \ref{sec5} we concentrate on the qualitative similarities between our
model and the results of numerical simulations by RUKL02 on the disk
structure. Using their numerical data, we roughly estimate the radius
where the disk structure is significantly changed and magnetic braking is
efficient. We summarize and discuss our results in \S\ \ref{sec6}.

\section{BASIC EQUATIONS}\label{sec2}

The steady-state equation of conservation of angular momentum for an
axisymmetric magnetized disk around a neutron star having a dipole moment
aligned with its rotation axis can be written in cylindrical coordinates 
($r$, $\phi $, $z$) as
\begin{eqnarray}
r\rho v_{r}\frac{\partial }{\partial r}\left( rv_{\phi }\right) +r\rho v_{z}
\frac{\partial }{\partial z}\left( rv_{\phi }\right)&=&\frac{\partial }{
\partial r}\left[ \nu \rho r^{3}\frac{\partial }{\partial r}\left( \frac{
v_{\phi }}{r}\right) \right] \nonumber \\
&&+\frac{rB_{r}}{4\pi }\frac{\partial }{\partial r}\left( rB_{\phi }\right) \nonumber \\
&&+\frac{r^{2}B_{z}}{4\pi }\frac{\partial B_{\phi }}{\partial z}\;,  \label{angmom}
\end{eqnarray}
where $\nu $ is the coefficient of viscosity. The conservation of mass for a
steady-state, axisymmetric disk is given by the continuity equation,
\begin{equation}
\frac{\partial }{\partial r}\left( r\rho v_{r}\right) +\frac{\partial }{
\partial z}\left( r\rho v_{z}\right) =0\;,  \label{contin}
\end{equation}
which can be vertically integrated to yield the constant mass influx
condition,
\begin{equation}
-2\pi \int_{-H}^{H}r\rho v_{r}\,dz=\dot{M}\;,  \label{mdot}
\end{equation}
provided that $\rho v_{z}$ is negligible at $z=\pm H$, a valid assumption
except for the case of strong winds or magnetically driven outflows. Here, 
$\dot{M}$ is the constant mass inflow rate and $H$ is the half-thickness of
the disk.

In a geometrically thin accretion disk (i.e., $H\ll r$), the angular
momentum balance (\ref{angmom}) can be approximated as
\begin{eqnarray}
\frac{\partial }{\partial r}\left( r\rho v_{r}r^{2}\Omega \right) +\frac{
\partial }{\partial z}\left( r\rho v_{z}r^{2}\Omega \right) &=&\frac{\partial 
}{\partial r}\left( \nu \rho r^{3}\frac{\partial \Omega }{\partial r}\right) \nonumber \\
&&+\frac{r^{2}B_{z}}{4\pi }\frac{\partial B_{\phi }}{\partial z}\;,  \label{agmom2}
\end{eqnarray}
where $v_{\phi }=\Omega r$ is used and the $r\phi $ component of the
magnetic stress is neglected in comparison with the last term in equation 
(\ref{angmom}), for the ratio of their magnitudes in the thin disk limit gives
\begin{equation}
\frac{\left\vert rB_{r}\partial _{r}\left( rB_{\phi }\right) \right\vert }{
\left\vert r^{2}B_{z}\partial _{z}B_{\phi }\right\vert }\sim \left\vert 
\frac{B_{r}}{B_{z}}\right\vert \left( \frac{H}{r}\right) \ll 1\;,  \label{comp}
\end{equation}
provided $\left\vert B_{r}/B_{z}\right\vert \ll r/H$, a condition that can
only be violated for extremely large disk conductivities.

The vertically averaged form of the angular momentum equation 
(\ref{agmom2}), together with the use of equation (\ref{mdot}), is finally
obtained as
\begin{equation}
\frac{d}{dr}\left( 2\pi \nu \Sigma r^{3}\frac{d\Omega }{dr}+\dot{M}
r^{2}\Omega \right) =-r^{2}B_{z}B_{\phi }^{+}\;,  \label{agmom3}
\end{equation}
where $B_{\phi }^{+}(r)\equiv B_{\phi }(r,z=H)=-B_{\phi }(r,z=-H)$ due to
the field antisymmetry and $\Sigma $ is the surface density function defined
by
\begin{equation}
\Sigma (r)=\int_{-H}^{H}\rho (r,z)\,dz=2\rho H\;,  \label{sigma}
\end{equation}
where $\rho $ denotes simply the vertically averaged density. The last term
in equation (\ref{angmom}) was also identified as the dominant magnetic term
in the early derivation of equation (\ref{agmom3}) by \citet{Camp92} for
steady disks and by \citet{LRN94} in a form appropriate for time-dependent
disks with outbursts.

The vertical component of the magnetic field threading the disk can be
expressed, in general, as
\begin{equation}
B_{z}(r)=-s(r)\mu _{\ast }r^{-3}\;,  \label{dipole}
\end{equation}
where $\mu _{\ast }=\left( 1/2\right) B_{\ast }R_{\ast }^{3}$ is the dipole
moment of the neutron star of radius $R_{\ast }$ in terms of its polar
surface field strength $B_{\ast }$ and $s(r)$ represents a screening
coefficient which accounts for the effect of induced currents on the pure
stellar dipole field in a partially diamagnetic disk. In a self-consistent
treatment, the radial dependence of $s$ must be properly found from a
detailed analysis of the magnetospheric current system in the corona. The
modelling of such a disk-corona-star system is beyond the scope of the
present study. In this work, we will rather assume that the vertical
component of the magnetic field threading the disk can be represented by
\begin{equation}
B_{z}(r)=-s_{\mathrm{eff}}\,\frac{\mu _{\ast }}{r^{3}}\;,  \label{scrdip}
\end{equation}
where $s_{\mathrm{eff}}$ is an effective screening coefficient taken to be a
constant over a wide range of radii (Wang 1995).

The azimuthal component of the magnetic field at the surface of a disk in
vertical hydrostatic equilibrium (i.e., $v_{z}=0$) can be estimated from the
toroidal component of the induction equation, which can be written under the
assumptions of steady-state and axisymmetry as
\begin{eqnarray}
\frac{\partial }{\partial z}\left( \Omega rB_{z}\right) &+& \frac{\partial }{
\partial r}\left( \Omega rB_{r}\right) -\frac{\partial }{\partial r}\left(
v_{r}B_{\phi }\right) \nonumber \\
&+&\frac{\partial }{\partial r}\left( \eta \frac{
\partial B_{\phi }}{\partial r}\right) +\frac{\partial }{\partial z}\left(
\eta \frac{\partial B_{\phi }}{\partial z}\right) =0\;,  \label{torind}
\end{eqnarray}
where $\eta $ is the magnetic diffusivity of the disk.
Assuming that all quantities change on a length scale $r$ in the radial
direction and $H$ in the vertical direction, the significance of the second
term compared with the first term in equation (\ref{torind}) can be
estimated as
\begin{equation}
\frac{\left\vert \partial _{r}\left( \Omega rB_{r}\right) \right\vert }{
\left\vert \partial _{z}\left( \Omega rB_{z}\right) \right\vert }\sim
\left\vert \frac{B_{r}}{B_{z}}\right\vert \left( \frac{H}{r}\right) \ll 1\;,  \label{comp1}
\end{equation}
provided we have $\left\vert B_{r}\right\vert \lesssim \left\vert
B_{z}\right\vert $. The last term in equation (\ref{torind}) dominates over
the fourth and third terms:
\begin{equation}
\frac{\left\vert \partial _{r}\left( \eta \partial _{r}B_{\phi }\right)
\right\vert }{\left\vert \partial _{z}\left( \eta \partial _{z}B_{\phi
}\right) \right\vert }\sim \left( \frac{H}{r}\right) ^{2}\ll 1  \label{comp2}
\end{equation}
and
\begin{equation}
\frac{\left\vert \partial _{r}\left( v_{r}B_{\phi }\right) \right\vert }{
\left\vert \partial _{z}\left( \eta \partial _{z}B_{\phi }\right)
\right\vert }\sim \frac{\left\vert v_{r}\right\vert H}{\eta }\left( \frac{H}{
r}\right) \ll 1\;,  \label{comp3}
\end{equation}
provided $\eta \gtrsim \left\vert v_{r}\right\vert H$ is satisfied. The
consistency of equation (\ref{comp1}) with equation (\ref{comp3}) can be
justified using the poloidal component of the induction equation which can
be written as
\begin{equation}
\frac{\partial B_{r}}{\partial z}-\frac{\partial B_{z}}{\partial r}+\frac{
v_{r}B_{z}}{\eta }=0\;,  \label{polind}
\end{equation}
where the disk is assumed to be in vertical hydrostatic equilibrium, i.e.,
$v_{z}=0$. For very small radial inflow velocities, i.e., $v_{r}\ll \eta /r$,
equation (\ref{polind}) yields $\left\vert B_{r}\right\vert \sim \left(
H/r\right) \left\vert B_{z}\right\vert $ and the distortion of the external
field due to accretion becomes negligible. In this case the estimates in
equations (\ref{comp1}) and (\ref{comp2}) are both $O((H/r)^{2})\ll 1$,
while the estimate in (\ref{comp3}) is smaller than $O((H/r)^{2})$ by the
factor $v_{r}r/\eta $. If, on the other hand, $\left\vert B_{r}\right\vert
\lesssim \left\vert B_{z}\right\vert $, as is likely when the field is
distorted, then $\left\vert \partial _{r}B_{z}\right\vert \ll \left\vert
\partial _{z}B_{r}\right\vert $ and we obtain
\begin{equation}
B_{r}^{+}=-\frac{v_{r}H}{\eta }B_{z}  \label{radB}
\end{equation}
from the vertical integration of equation (\ref{polind}), where 
$B_{r}^{+}(r)\equiv B_{r}(r,z=H)=-B_{r}(r,z=-H)$ following the field
antisymmetry (see Lovelace, Romanova, \& Newman 1994; also LRBK95).
When equation (\ref{radB}) holds, approximations (\ref{comp1})
and (\ref{comp3}) are both $\lesssim O(H/r)\ll 1$.

Taking the conditions (\ref{comp1}), (\ref{comp2}), and 
(\ref{comp3}) to be satisfied, the toroidal component of the induction equation 
(\ref{torind}) can be safely reduced to
\begin{equation}
rB_{z}\frac{\partial \Omega }{\partial z}+\frac{\partial }{\partial z}\left(
\eta \frac{\partial B_{\phi }}{\partial z}\right) =0\;.  \label{torind1}
\end{equation}
The vertical gradient of angular velocity depends on the scale height
over which the plasma is brought into corotation with the neutron star. In
the case of a force-free magnetosphere outside the disk, this scale height
can be estimated as $H$, the half-thickness of the disk, since the magnetic
energy density becomes much greater than the kinetic energy density of the
plasma for $\left\vert z\right\vert \geq H$. The resulting twisting of
field loops due to this huge vertical shear gradient may lead to an open field
line configuration unless $\left\vert \partial \Omega /\partial z\right\vert
<\left\vert \Omega _{\ast }-\Omega \right\vert /H$. The recent
time-dependent simulations by RUKL02 have shown that the field line opening is
strongly suppressed for a relatively dense corona outside the disk. Assuming
that the vertical shear along a flux tube linking the star and the disk is
reduced by the inertial effect of a corona, the integration of equation 
(\ref{torind1}) over $z$ can be approximated as
\begin{equation}
B_{\phi }^{+}=\epsilon r\left( \Omega _{\ast }-\Omega \right)H\eta
^{-1}B_{z}\;,  \label{aziB}
\end{equation}
where the transition from the angular velocity of the disk plasma, $\Omega $, 
to the rotational rate of the neutron star, $\Omega _{\ast }$, takes place
within an effective scale height of $\epsilon ^{-1}H$ with $\epsilon<1$
(see Campbell 1992; also LRBK95 for other derivations of eq. [\ref{aziB}]).
Here, $\epsilon $ is a shear reduction factor which may arise from a differentially
rotating corona to result in a small field twist at the surface of the disk. Even if
the corona is nearly force-free, it cannot be current-free. As the magnetic torque
is transmitted to the neutron star by the poloidal currents flowing across the disk
surface through the magnetosphere, the rigid-body rotation of the corona may not
be realized (see W87; also CH98 for a similar shear reduction factor).

We consider a specific disk region (e.g., a narrow boundary layer of
width $\delta r\lesssim H\ll r$) where the contribution from the radial
gradient terms in equation (\ref{torind}) is not negligible. In this case,
the angular velocity transition from $\Omega $ to $\Omega _{\ast }$ occurs
within $\delta r$. The approximate balance between the second term (field
generating) and the third (advection) and fourth (dissipation) terms in
equation (\ref{torind}) yields a crude estimation for the toroidal field
component,
\begin{equation}
B_{\phi }^{+}\approx r\left( \Omega _{\ast }-\Omega \right) \left( \delta
r\right) \eta ^{-1}B_{r}^{+}\sim r\left( \Omega _{\ast }-\Omega \right)
H\eta ^{-1}B_{z}\;,  \label{torB}
\end{equation}
provided the width of the boundary layer is assumed to be the
electromagnetic screening length for the poloidal magnetic field, i.e. 
$\delta r\simeq \eta /v_{r}$ (see GL79). The vertical gradient terms are not
taken into account for simplicity in the radial integration of equation 
(\ref{torind}). The last step in (\ref{torB}) follows from $B_{r}^{+}/B_{z}\sim
H/\delta r$ according to equation (\ref{polind}). This analysis indicates
that the form of the expression for the azimuthal field component in
equation (\ref{aziB}) may generally be applicable for the entire disk
regardless of the radial extension of the angular velocity transition zone.

In the thin disk limit, the vertically integrated forms of the $r$ and $z$
components of the momentum equation can be written as
\begin{equation}
\left( \Omega _{\mathrm{K}}^{2}-\Omega ^{2}\right) rH-\frac{B_{z}B_{r}^{+}}{
4\pi \rho }\simeq 0\;,  \label{radmom}
\end{equation}
and
\begin{equation}
c_{s}^{2}\simeq \frac{1}{2}\Omega _{\mathrm{K}}^{2}H^{2}+\frac{(B_{\phi
}^{+})^{2}+(B_{r}^{+})^{2}}{8\pi \rho }\;,  \label{VHE}
\end{equation}
where $c_{s}=(P/\rho )^{1/2}$ is the speed of sound in terms of the
mid-plane temperature of the disk, and $\Omega _{\mathrm{K}}$ is the
Keplerian angular velocity given by
\begin{equation}
\Omega _{\mathrm{K}}(r)=\left( \frac{GM_{\ast }}{r^{3}}\right) ^{1/2}\;,  \label{kepler}
\end{equation}
where $M_{\ast }$ is the mass of the neutron star. The magnetic tension
force in equation (\ref{radmom}) is the main agent responsible for the
deviation of $\Omega (r)$ from $\Omega _{\mathrm{K}}(r)$ if the accretion
disk is thin and the poloidal magnetic field varies on a length scale $r$.
The vertical equilibrium of a magnetically threaded thin disk is given by
equation (\ref{VHE}), where the gas pressure is balanced by the magnetic
pressure of the horizontal field components in addition to the vertical
component of gravity. Using equations (\ref{radmom}) and (\ref{VHE}), it is
possible to make an order of magnitude estimate for the sound speed. In the
absence of a strong global magnetic field, the rotational profile is nearly
Keplerian and $c_{s}\simeq \Omega _{\mathrm{K}}H$. The rotation of the disk
matter at a sub-Keplerian rate can only be realized provided we have
\begin{equation}
\frac{\left\vert B_{r}^{+}\right\vert B}{4\pi \rho }\lesssim \Omega _{
\mathrm{K}}^{2}Hr  \label{magbal}
\end{equation}
in equation (\ref{radmom}) with $\left\vert B_{z}\right\vert \sim B$. If the
magnetic diffusivity is turbulent in nature, then the radial component of
the magnetic field can be significant, i.e., $\left\vert
B_{r}^{+}\right\vert \lesssim B$, and the vertical equilibrium (\ref{VHE})
is satisfied by the balance between the gas and magnetic pressures alone,
\begin{equation}
c_{s}^{2}\approx \frac{B^{2}}{4\pi \rho }\lesssim \Omega _{\mathrm{K}}^{2}Hr\;,  \label{sound}
\end{equation}
since the gravity term is relatively small in a thin disk, i.e., $\Omega _{
\mathrm{K}}^{2}H^{2}\ll \Omega _{\mathrm{K}}^{2}Hr$.

After equation (\ref{sound}), we propose the following prescription for
the sound speed:
\begin{equation}
c_{s}^{2}=\xi (r)\,\Omega _{\mathrm{K}}^{2}Hr\;,  \label{soundpresc}
\end{equation}
where $\xi (r)<1$ is a dimensionless factor of unknown radial dependence.
Its explicit form can only be deduced from a detailed analysis of the field configuration
and density distribution in the disk.

In this work, we assume that the magnetic diffusivity of the disk is of the
same physical origin (i.e., turbulence) as the viscosity, and we adopt the
original $\alpha $ formalism \citep{SS73} to write
\begin{equation}
\eta =\alpha _{d}\,\frac{c_{s}^{2}}{\Omega _{\mathrm{K}}}\;,  \label{mdiff}
\end{equation}
where $\alpha _{d}\lesssim 1$ is a dimensionless numerical factor.

Combining equations (\ref{soundpresc}) and (\ref{mdiff}), the azimuthal field
component at the disk surface (see eq. [\ref{aziB}]) is simplified to
\begin{equation}
B_{\phi }^{+}=\gamma _{\phi }\left( \Omega _{\ast }-\Omega \right) \Omega _{
\mathrm{K}}^{-1}B_{z}\;,  \label{aziB1}
\end{equation}
where $\gamma _{\phi }\equiv \epsilon /\xi \alpha _{d}$ will be treated as a
constant for simplicity (see, e.g., Livio \& Pringle 1992). As we will see in
\S\ \ref{sec3}, the vertically integrated dynamical viscosity, $\nu \Sigma $,
for a magnetically threaded disk changes with distance. The radial variation
of $\nu \Sigma $ can be achieved in general provided both the coefficient
of viscosity $\nu $ and the surface density $\Sigma $ are functions of radius.
As the viscosity $\nu $ and diffusivity $\eta $ have the same, turbulent
nature by assumption, the magnetic Prandtl number, $\nu /\eta$, should
be of order one at different radii of the disk. This also requires that the
parameter $\xi$ should have a radial dependence in general
for self-consistency. In this case, $\gamma _{\phi }$ would not be
approximated as a constant unless the shear reduction factor $\epsilon$
changes with distance in more or less the same way as $\xi$ does.
Our physical motivation for taking $\gamma _{\phi }$ as a constant is the
following. The compressive effect of the magnetic field on the vertical equilibrium of the
disk becomes negligible in the weakly magnetized outer disk regions. The gas pressure
is mainly balanced by the gravitational force and $c_{s}\simeq \Omega _{\mathrm{K}}H$
(see eq. [\ref{VHE}]). According to equation (\ref{soundpresc}), this is equivalent to
choose $\xi (r)\approx H/r$ in the outer disk. The resulting toroidal field is
$O(r/H)B_{z}\gg B_{z}$ unless $\epsilon (r)\approx H/r$. The shear reduction factor
$\epsilon$ is therefore expected to decrease from $O(1)$ at small radii to $O(H/r)$ at large
radii provided the disk is threaded by the stellar field lines that are closed and stable
at least to some extent beyond the corotation radius. As there appears to be a
correlation between $\epsilon (r)$ and $\xi (r)$, we approximate their ratio as a constant
throughout the entire disk.

In the absence of a complete theory of the disk-star magnetic interaction,
we will not attempt to obtain a full disk solution at this stage; we will
rather concentrate on the conservation of angular momentum to treat the
dynamics of disk accretion under the action of an external magnetic field of
stellar origin. Using equation (\ref{aziB1}), it follows from 
equation~(\ref{agmom3}) that
\begin{equation}
\frac{d}{dr}\left( 2\pi \nu \Sigma r^{3}\frac{d\Omega }{dr}+\dot{M}
r^{2}\Omega \right) =-\gamma _{\phi }r^{2}\left( \Omega _{\ast }-\Omega
\right) \Omega _{\mathrm{K}}^{-1}B_{z}^{2}  \label{agmom4}
\end{equation}
which can be solved for the rotation profile, $\Omega (r)$, for a given
vertically integrated dynamical viscosity, $\nu \Sigma $, if $B_{z}(r)$ is
known.

\section{DYNAMICAL VISCOSITY}\label{sec3}

In the following, we will deal with the angular momentum balance (see \S\ ~\ref{sec2}) 
to derive an expression for the vertically integrated dynamical
viscosity. Before attempting to solve equation (\ref{agmom4}), we scale the
variable quantities $r$, $\nu \Sigma $, and $\Omega $ by their typical
values such that $x\equiv r/r_{\mathrm{in}}$ is a dimensionless coordinate
in units of the inner disk radius $r_{\mathrm{in}}$ where we assume $\Omega
=\Omega _{\ast }$, $f(x)\equiv 3\pi \nu \Sigma /\dot{M}$ is a dimensionless
function for the dynamical viscosity, and 
$\omega (x)\equiv \Omega /\Omega _{\mathrm{K}}(r_{\mathrm{in}})$ is a 
dimensionless angular velocity for the
disk plasma. Finally, we use equation (\ref{scrdip}) for the screened dipole
field to rewrite equation (\ref{agmom4}) in a non-dimensional form:
\begin{equation}
\frac{d}{dx}\left[ \frac{2}{3}\,f(x)\,x^{3}\frac{d\omega }{dx}+x^{2}\omega 
\right] =-\beta \left( \omega _{\ast }-\omega \right) x^{-5/2}\;,  \label{ndam}
\end{equation}
where $\omega _{\ast }\equiv \Omega _{\ast }/\Omega _{\mathrm{K}}(r_{\mathrm{
in}})=(r_{\mathrm{co}}/r_{\mathrm{in}})^{-3/2}$ with the corotation radius 
$r_{\mathrm{co}}=(GM_{\ast }/\Omega _{\ast }^{2})^{1/3}$ and $\beta \equiv
\gamma _{\phi }\,s_{\mathrm{eff}}^{2}(r_{\mathrm{A}}/r_{\mathrm{in}})^{7/2}$
with the Alfv\'{e}n radius given by
\begin{eqnarray}
r_{\mathrm{A}}&=&\dot{M}^{-2/7}\mu _{\ast }^{4/7}\left( GM_{\ast }\right)
^{-1/7}  \nonumber \\
&\simeq & 3.4\times 10^{8}\;\mathrm{cm\;}\dot{M}_{17}^{-2/7}\mu _{\ast
30}^{4/7}\left( \frac{M_{\ast }}{1.4\;M_{\odot }}\right) ^{-1/7}\;.  \label{Alfv}
\end{eqnarray}
Here, $\dot{M}_{17}$ is the mass accretion rate expressed in units of 
$10^{17}$ g s$^{-1}$ and $\mu _{\ast 30}$ is the magnetic dipole moment in
units of $10^{30}$ G cm$^{3}$. If the neutron star magnetic field is
sufficiently weak for a given mass accretion rate, e.g., $B_{\ast }\lesssim
10^{7}$ G in equation (\ref{Alfv}), then we have $r_{\mathrm{A}}<R_{\ast }$
and the disk may extend down to the surface of the central object (i.e., $r_{
\mathrm{in}}=R_{\ast }$). For such weakly magnetized systems also known as
the standard $\alpha $-disks (SS), we get $\beta \simeq 0$. The same is true
also if the disk plasma is perfectly diamagnetic. In the conventional
picture of $\alpha $-disks, the rotation curve is Keplerian throughout the
accretion flow except in a narrow boundary layer situated at the inner edge
of the disk where the angular velocity of the plasma changes from $\Omega _{
\mathrm{K}}(R_{\ast })$ to $\Omega _{\ast }$. In a standard thin disk around
a non-magnetic star, we can integrate equation (\ref{ndam}) with $\beta =0$
to find $f(x)=1-\zeta x^{-1/2}$ using the Keplerian profile, $\omega
(x)=x^{-3/2}$. The integration constant $\zeta $, here, denotes the inflow
rate of angular momentum $\dot{J}$, across a cylindrical boundary at radius 
$r$, in terms of the angular momentum flux carried by matter onto the neutron
star through the inner edge of the disk, i.e., $\dot{J}=\zeta \dot{M}R_{\ast
}^{2}\Omega _{\mathrm{K}}(R_{\ast })$. According to the standard approach
\citep{LBP74}, there is no viscous torque acting on the inner
disk. This corresponds to $\zeta \simeq 1$ and the boundary layer is
extremely thin in extension. However, the angular velocity transition region
is not necessarily narrow as Glatzel showed \citep{Glat92}
if $\zeta $ is treated as a free parameter in $f(x)$, which in turn can be
used to solve for the structure of the accretion flow \citep{Fuji95}. 
In the present investigation, we will derive a form of vertically
integrated dynamical viscosity in a way similar to the work of G92 but
generalized to treat the disk interacting with a magnetic star. In our case, 
$\beta \neq 0$ and the angular momentum exchange between the magnetosphere
and the disk is not negligible except for sufficiently large distances from
the magnetically braked inner disk regions. The Keplerian rotation can only
be regarded as a particular solution that can be employed to obtain a
prescription for the dynamical viscosity, which may hold throughout the
accretion flow provided the angular motion of the inner disk plasma can be
matched to the rotation rate of the stellar magnetosphere. The radial
integration of equation (\ref{ndam}) for a Keplerian angular velocity
profile yields
\begin{equation}
f(x;\beta ,j,\omega _{\ast })=1-\frac{2}{3}\beta \omega _{\ast }x^{-2}+\frac{
1}{3}\beta x^{-7/2}-jx^{-1/2}\;,  \label{dvis}
\end{equation}
where $j\equiv \dot{J}/\dot{M}r_{\mathrm{in}}^{2}\Omega _{\mathrm{K}}(r_{
\mathrm{in}})$ is the net angular momentum flux into the neutron star unless
there are some other efficient mechanisms of angular momentum loss for the
disk material such as winds associated with the opening of field lines.
Although $j$, here, appears to be an arbitrary integration constant, it is
the dimensionless torque applied by the disk on the star, i.e., the star
spins up (or spins down) for $j>0$ (or $j<0$).

A vertically integrated dynamical viscosity form similar to the one we
consider in equation (\ref{dvis}) was derived and used previously by
\citet{BC98} to solve for the disk structure around
strongly magnetic accretors. These authors, however, assumed $j=0$ to ensure
that the disk structure matches that of weakly magnetized accretion at a
sufficiently large radius $r_{\mathrm{out}}$. Adopting $\nu \Sigma =\dot{M}
/3\pi $, that is, $f(x=r_{\mathrm{out}}/r_{\mathrm{in}})=1$ as an arbitrary
outer boundary condition, they integrated the relevant MHD equations in the
radial direction from the outer disk regions inwards. Note that $f=1$ at
large distances whether $j=0$ or not. However, the structure at finite $x$
depends on the relative importance of the terms, and the value of $j$ will
be an important parameter in determining the structure. BC98 did not
attempt to fit their solution to a magnetically braked inner accretion flow;
they assumed that the Keplerian rotation holds throughout the disk.
Their model implies that the torque on the star is always zero, i.e., $j=0$,
and the angular velocity of the plasma still remains Keplerian in the inner
parts of the disk even though the viscous dissipation there due to
magnetically enhanced viscous stress (see the third term on the r.h.s. of
eq. [\ref{dvis}]) leads to the divergence of the disk height as a result of
elevated temperature. Both assumptions are very restrictive and non-physical
in the sense that the choice of $j$ becomes quite significant at small radii
although it seems to have no effect on the disk structure at large radii
(see \S\ \ref{sec4}) and the huge viscous dissipation estimated for the
inner disk regions results from using overestimated angular frequencies
(e.g., Keplerian frequencies) together with a magnetically enhanced
dynamical viscosity. The last argument can be made clear if we consider the
conservation of energy for a radiatively efficient disk,
\begin{equation}
2\sigma T_{s}^{4}\approx \int_{-H}^{H}Q_{v}dz
=\frac{GM_{\ast }\dot{M}}{3\pi r_{\mathrm{in}}^{3}}
\left( x\frac{d\omega }{dx}\right) ^{2}f(x;\beta ,j,\omega _{\ast })\;,  \label{visd}
\end{equation}
where $T_{s}$ is the effective surface temperature of the disk and $
Q_{v}=\nu \rho (r\partial _{r}\Omega )^{2}$ is the viscous dissipation rate
per unit volume. The approximation sign in equation (\ref{visd}) indicates
that there may exist additional sources of heat (e.g., Ohmic dissipation)
other than viscosity in the disk. If the Keplerian assumption, i.e., $\omega
(x)=x^{-3/2}$, were valid throughout the entire disk as in the work of BC98,
then we would certainly find huge temperatures as $x\rightarrow 1$ for
sufficiently high $\beta $ and low $\omega _{\ast }$ values (see, e.g.,
CH98 for the corresponding values) in the case of $j=0$ (see also, eq.
[\ref{dvis}]).

What we expect to find in steady accretion is that the inner boundary
condition, that is, the fastness of the stellar magnetosphere $\omega _{\ast
}$, adjusts the rotational disk dynamics for a given torque $j$. The quantity
$\beta $ in equation (\ref{dvis}), which reflects the magnetic configuration in the
inner disk, is not an independent parameter. Rather, it is linked intimately
to $j$ and its value can be calculated (see \S\ \ref{sec4}) when the angular
velocity profile is determined from equation (\ref{ndam}).

The definition of $\beta $ with $r_{\mathrm{in}}=\omega _{\ast }^{2/3}r_{
\mathrm{co}}$, yields for a typical X-ray binary pulsar:
\begin{equation}
\beta \simeq 12\;\gamma _{\phi }\,s_{\mathrm{eff}}^{2}\,\omega _{\ast
}^{-7/3}\dot{M}_{17}^{-1}P_{\ast }^{-7/3}\mu _{\ast 30}^{2}\left( \frac{
M_{\ast }}{1.4\;M_{\odot }}\right) ^{-5/3}\;,  \label{beta}
\end{equation}
where $P_{\ast }$ is the spin period of the neutron star. Using the torque
expression, $N_{\ast }=I_{\ast }\dot{\Omega}_{\ast }=j\dot{M}(GM_{\ast }r_{\mathrm{in}
})^{1/2}$ with a moment of inertia $I_{\ast }=\left( 2/5\right) M_{\ast
}R_{\ast }^{2}$, we find
\begin{eqnarray}
j &\simeq& -0.4\;\omega _{\ast }^{-1/3}\dot{M}_{17}^{-1}P_{\ast }^{-7/3}R_{\ast
6}^{2}\left( \frac{\dot{P}_{\ast }}{10^{-12}\,\mathrm{s\,s}^{-1}}\right) \nonumber \\
&&\times \left( \frac{M_{\ast }}{1.4\;M_{\odot }}\right) ^{1/3}\;,  \label{j}
\end{eqnarray}
where $R_{\ast 6}$ is the neutron star radius expressed in units of $10^{6}$
cm and $\dot{P}_{\ast }$ is the time rate of change of the spin period.

\begin{figure}
\epsscale{1.0} \plotone{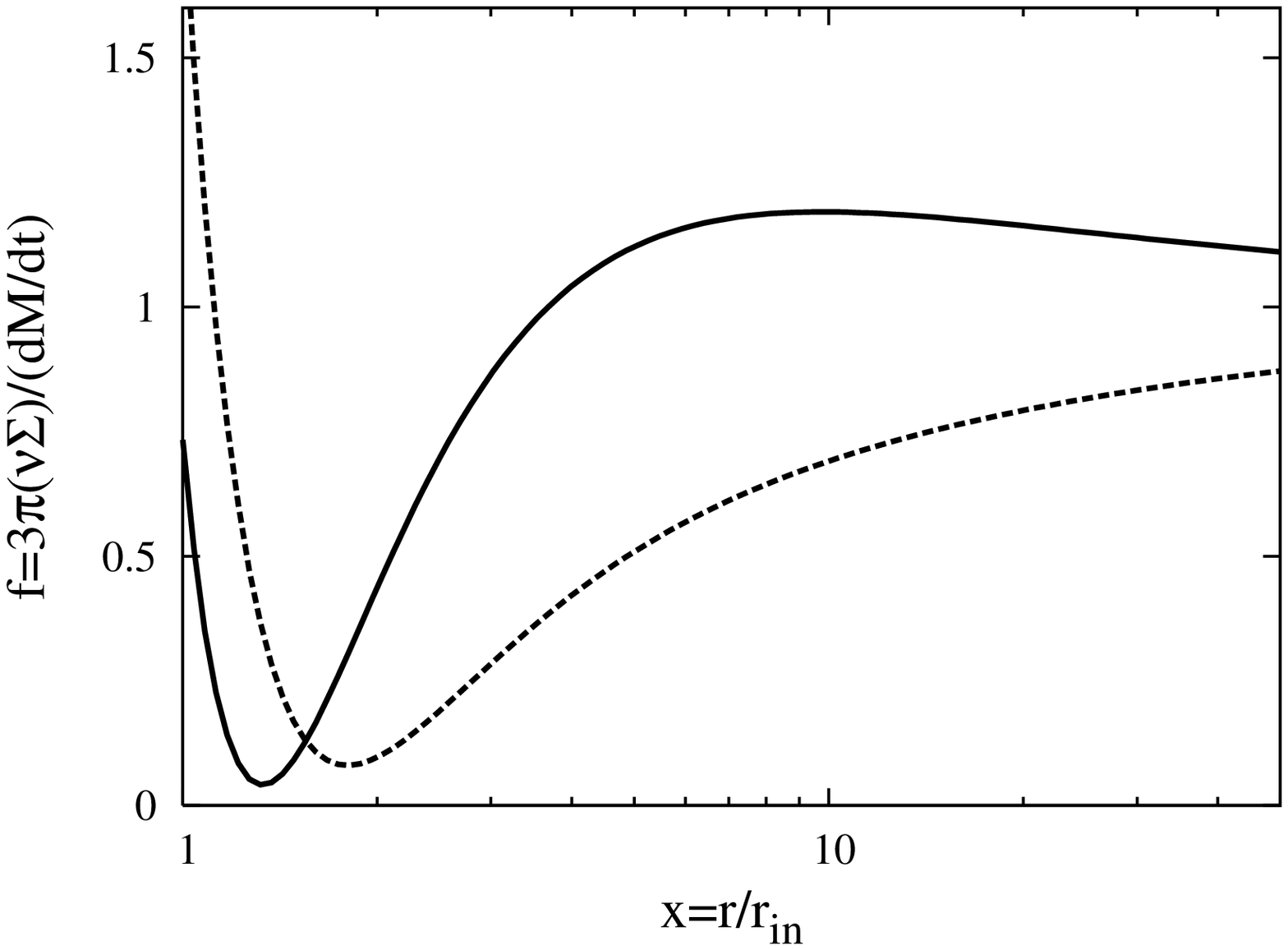} 
\caption{Radial variation of normalized vertically integrated dynamical viscosity $f(x)=3\pi \nu \Sigma /\dot{M}$
from the outer disk regions $x=50$ to the innermost radius $x=1$, where $x\equiv r/r_{\mathrm{in}}$, for different
values of parameters $(\beta ,j,\omega _{\ast })$. The
\emph{solid curve} corresponds to $f(x)$ with $\beta =16$, $\omega_{\ast }=0.6$, and $j=-0.8$ and the \emph{dashed curve}
to $f(x)$ with $\beta =13$, $\omega _{\ast }=0.3$, and $j=0.9$. \label{fig1}}
\end{figure}

We now examine some of the basic features of the dimensionless function we
introduced for the dynamical viscosity (see eq. [\ref{dvis}]) before we
present various types of rotation profiles in the next section, for the
behavior of $\omega (x)$ strongly depends on $f(x)$ and its first derivative 
$f^{\prime }(x)$ (see eq. [\ref{ndam}]). As an illustrative example of how
the dimensionless viscosity is affected by the last term in equation (\ref
{dvis}), we display in Figure~\ref{fig1} plots of $f(x)$ for two different values of
$j$. The dynamical viscosity function
shown by the solid curve in Figure~\ref{fig1} corresponds to $\beta =16$, $\omega
_{\ast }=0.6$, and $j=-0.8$. According to equations (\ref{beta}) and (\ref{j}
), this can be realized for a typical X-ray binary pulsar of spin period $1$
s if $\gamma _{\phi }\simeq 0.4$ with $s_{\mathrm{eff}}\simeq 1$ and $\dot{P}
_{\ast }\simeq 1.7\times 10^{-12}\,\mathrm{s\,s}^{-1}$. The dashed
curve (see Fig.~\ref{fig1}) with $\beta =13$, $\omega _{\ast }=0.3$, and $j=0.9$,
represents another physically plausible dynamical viscosity for the same
pulsar provided $\gamma _{\phi }\simeq 0.1$ for $s_{\mathrm{eff}}\simeq 0.8$
and $\dot{P}_{\ast }\simeq -1.5\times 10^{-12}\,\mathrm{s\,s}^{-1}$. Note
that the curve associated with a negative value of $j$ (spin-down) is
characterized by two local extrema, corresponding to a maximum and a
minimum, located respectively at $x_{2}$ and $x_{1}$, with $x_{1}<x_{2}$.
Also, note that there is no local maximum for $j>0$. The curves in Figure~\ref{fig1}
are plotted for illustrative values of $\beta $ and $\omega _{\ast }$. The
conclusions drawn from Figure~\ref{fig1} however can be generalized also for other
values of $\beta $ and $\omega _{\ast }$, if we consider the sign of the
second derivative of $f(x)$ at local extrema. The real extremum points, $x_{
\mathrm{ext}}=x_{1,2}$, can be found from $f^{\prime }(x_{\mathrm{ext}})=0$
as
\begin{equation}
x_{1,2}=\frac{1}{9j^{2}}\left[ \left( -36\beta
\omega _{\ast }\pm 9\sqrt{16\beta ^{2}\omega _{\ast }^{2}+21\beta j}\right)
j^{2}\right] ^{2/3}\;,  \label{expt}
\end{equation}
provided we have $16\beta \omega _{\ast }^{2}+21j\geq 0$. The first root,
$x_{1}$, specified by the plus sign within the parenthesis in equation (\ref
{expt}) can be identified to be the local minimum, whereas the second root,
$x_{2}$, with minus sign represents the local maximum. Equation (\ref{expt})
then, implies that $x_{1}<x_{2}$. The local minima of both curves in
Figure~\ref{fig1} are inside the corotation radii, i.e., $x_{1}\simeq 1.3<
x_{\mathrm{co}}\simeq 1.4<x_{2}\simeq 9.8$ for $\omega _{\ast }=0.6$ and $
x_{1}\simeq 1.8<x_{\mathrm{co}}\simeq 2.2$ for $\omega _{\ast }=0.3$, where
$x_{\mathrm{co}}\equiv r_{\mathrm{co}}/r_{\mathrm{in}}$.
As we will see in \S\ \ref{sec4}, $x_{1}<r_{\mathrm{co}}/r_{\mathrm{in}
}$ is always satisfied if the rotation of the disk matter is nearly Keplerian
for $r\geq r_{\mathrm{co}}$. The plasma is progressively brought
into corotation with the neutron star in an inner transition zone of radial
extension $\delta r_{\mathrm{in}}$. The angular velocity of the inner disk
matter starts deviating from its Keplerian value at $r\lesssim x_{1}r_{\mathrm{in}}$,
for the magnetic stress dominates over the shear stress when the dynamical
viscosity becomes sufficiently small, that is, when $f(x_{1})\ll 1$. The width
of the inner transition zone can be then estimated as $\delta \lesssim x_{1}-1$.

\section{APPLICATION TO OBSERVED BINARY X-RAY PULSARS}\label{sec4}

\begin{figure}
\epsscale{1.0} \plottwo{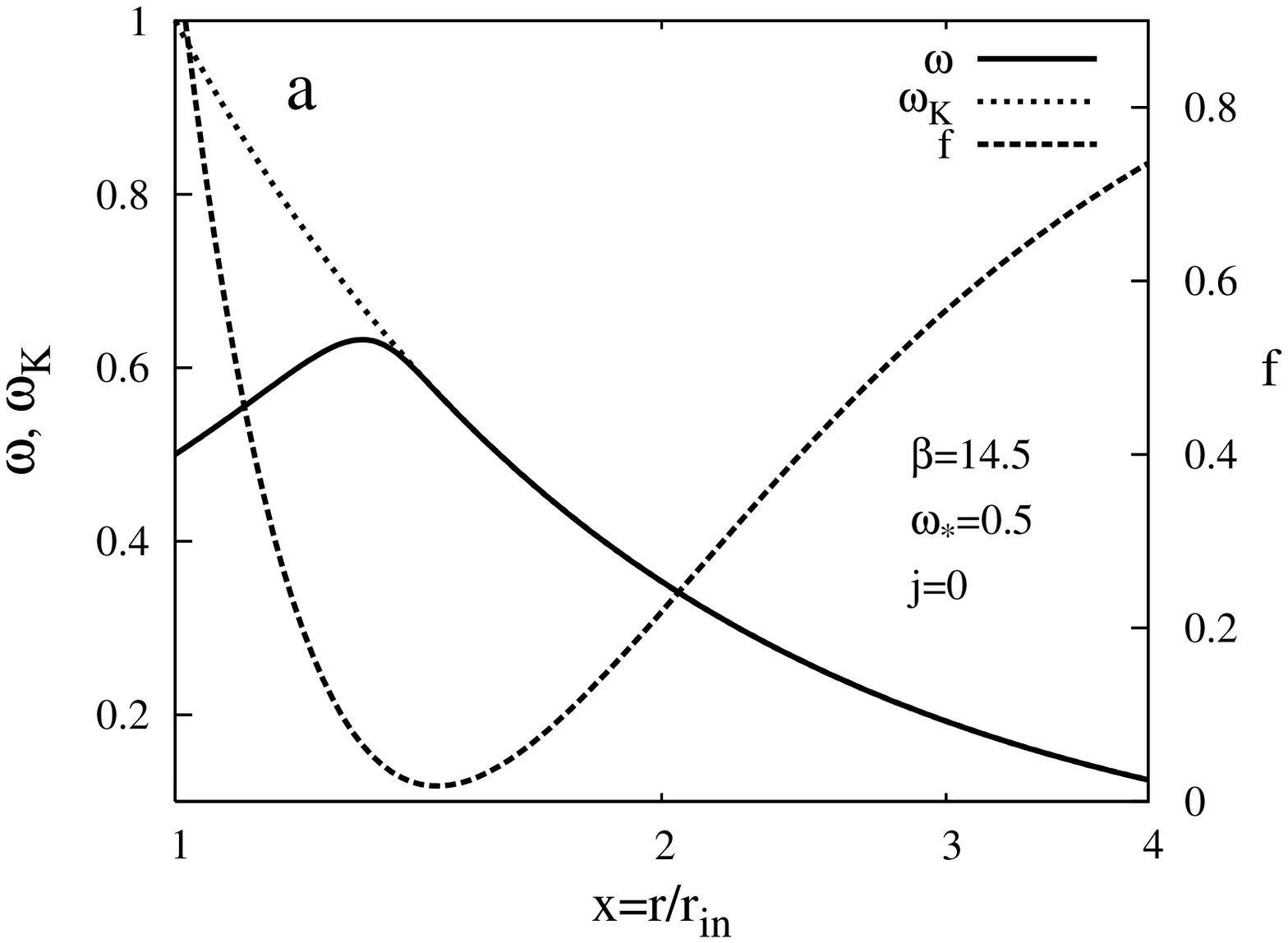}{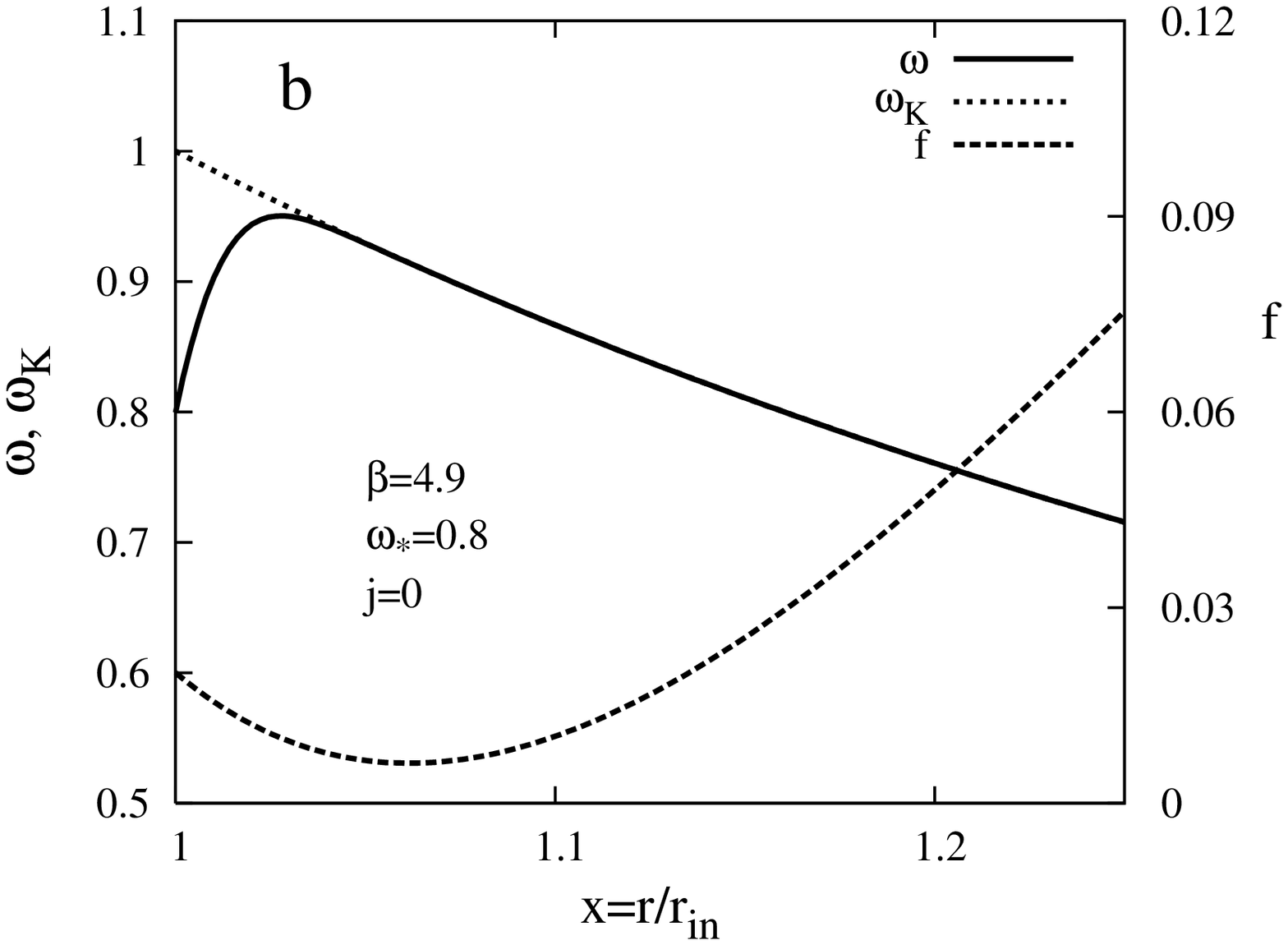} 
\caption{Radial variation of angular velocity $\omega (x)\equiv \Omega /\Omega _{\mathrm{K}}(r_{\mathrm{in}})$ and
the corresponding normalized vertically integrated dynamical viscosity $f(x)=3\pi \nu \Sigma /\dot{M}$ for a disk around
a magnetized star in spin equilibrium. The rotation rate profiles $\omega (x)$ shown by \emph{solid curves} in panels
(a) and (b) obtain for $\beta =14.5$, $\omega_{\ast }=0.5$, $j=0$ and $\beta =4.9$, $\omega_{\ast }=0.8$, $j=0$,
respectively. In each panel the \emph{dashed curve} corresponds to $f(x)$ and the \emph{dotted curve} to the Keplerian
profile $\omega _{\mathrm{K}}(x)=x^{-3/2}$, where $x\equiv r/r_{\mathrm{in}}$. \label{fig2}}
\end{figure}

We solved the second order differential equation (\ref{ndam}) for $\omega
(x) $, substituting equation (\ref{dvis}) for the dynamical viscosity. As we
do not know a priori where the true inner disk radius is located, we tried
different values of $\omega _{\ast }$ for a given torque $j$. As an initial
guess for each set of ($\omega _{\ast }$, $j$), we calculated $\beta $ such
that $f(x_{1})=0$. We observed that the necessary boundary conditions, i.e., 
$\omega (x=1)=\omega _{\ast }$ and $\omega (x>x_{\mathrm{co}})\simeq
x^{-3/2} $, are satisfied to the desired accuracy in our numerical iteration only for
certain values of $\beta $ for which the dynamical viscosity nearly vanishes
at $x=x_{1}$. Our results for a rotator in spin equilibrium, i.e., $j=0$,
are shown in Figures~\ref{fig2}$a$ and 2$b$. As in Figure~\ref{fig1}, the corresponding
dynamical viscosities (dashed curves) and the angular velocity profiles
(solid curves) are plotted as functions of $x=r/r_{\mathrm{in}}$. The dotted curves
represent Keplerian rotation. Note that the angular velocity is braked by
the magnetic stress only inside $x_{\mathrm{co}}$, and the rotation of the
disk plasma is almost Keplerian outside $x_{\mathrm{co}}$. We calculate the
net torque applied by the disk on the star using a closed surface through
the corotation radius. The surface encloses the central object and excludes
part of the disk for $x\geq x_{\mathrm{co}}$. This choice avoids the
uncertainties involved in making torque calculations in the inner disk where 
$\Omega \neq \Omega _{\mathrm{K}}$. We write the torque as
\begin{equation}
N_{\ast }=N_{\mathrm{co}}^{\mathrm{mat}}+N_{\mathrm{co}}^{\mathrm{vis}}+N_{
\mathrm{mag}}\;,  \label{trq}
\end{equation}
where the angular momentum flux carried by the material stress through a
cylindrical surface area $4\pi r_{\mathrm{co}}H(r_{\mathrm{co}})$ is
\begin{equation}
N_{\mathrm{co}}^{\mathrm{mat}}=\dot{M}r_{\mathrm{co}}^{2}\Omega _{\mathrm{K}
}(r_{\mathrm{co}})\;.  \label{trqm}
\end{equation}
The contribution of the viscous stress to the flux of angular momentum
through the same surface is given by
\begin{eqnarray}
N_{\mathrm{co}}^{\mathrm{vis}}&=&\frac{\dot{M}}{3\pi }f(x_{\mathrm{co}})\,r_{
\mathrm{co}}^{2}\left( \frac{d\Omega _{\mathrm{K}}}{dr}\right) _{r_{\mathrm{
co}}}(2\pi r_{\mathrm{co}}) \nonumber \\
&=&-f(x_{\mathrm{co}})\dot{M}r_{\mathrm{co}
}^{2}\Omega _{\mathrm{K}}(r_{\mathrm{co}})\;.  \label{trqv}
\end{eqnarray}
The magnetic coupling outside the corotation radius yields a net spindown
torque,
\begin{equation}
N_{\mathrm{mag}}=-\int_{r_{\mathrm{co}}}^{\infty }r^{2}B_{z}B_{\phi
}^{+}\,dr=-\frac{1}{3}\beta x_{\mathrm{co}}^{-7/2}\dot{M}r_{\mathrm{co}
}^{2}\Omega _{\mathrm{K}}(r_{\mathrm{co}})  \label{trqmag}
\end{equation}
which, together with equations (\ref{trqm}) and (\ref{trqv}), gives $N_{\ast
}=j\dot{M}(GM_{\ast }r_{\mathrm{in}})^{1/2}$ according to equation (\ref
{dvis}). This justifies our interpretation of $j$ as the dimensionless
torque acting on the neutron star in units of the angular momentum flux
carried by matter through the inner edge of the disk. Also, note from
Figures~\ref{fig2}$a$ and 2$b$ that $x_{1}$ decreases (i.e., $x_{1}\rightarrow 1$)
while $\omega _{\ast }$ increases as expected. There is, however, an upper
limit for $\omega _{\ast }$ if the star is in rotational equilibrium since
\begin{equation}
\lim_{j\rightarrow 0}x_{1}(\beta ,j,\omega _{\ast })=\left( \frac{7}{8\omega
_{\ast \mathrm{,c}}}\right) ^{2/3}\geq 1\;,  \label{min}
\end{equation}
independent of $\beta $, following from equation (\ref{expt}). Thus, for
accreting systems in rotational equilibrium, $j=0$, the critical fastness
parameter $\omega _{\ast \mathrm{,c}}$ for the transition from spin-down to
spin-up depends on the angular velocity profile (width of the transition
zone) in equilibrium. Equation (\ref{min}) implies that the maximum possible
critical fastness parameter is $(\omega _{\ast \mathrm{,c}})_{\max }=0.875$,
a value quoted previously by \citet{Wang95} as the critical fastness
parameter for a star accreting from a turbulent disk. According to the
accretion torque model of W95, the Keplerian rotation holds throughout the
whole disk. Equivalently, in our model, this maximum critical fastness
parameter obtains if the angular velocity transition region is extremely
narrow or there is no transition region at all. This corresponds to the
$x_{1}=1$ limit among systems in spin equilibrium as seen from the comparison
of Figures~\ref{fig2}$a$ and 2$b$. The usual fastness parameter, $\omega _{\mathrm{s}
} $, can be expressed in terms of $\omega _{\ast }$ as
\begin{equation}
\omega _{\mathrm{s}}\equiv \frac{\Omega _{\ast }}{\Omega _{\mathrm{K}}(r_{0})
}=\left( \frac{r_{0}}{r_{\mathrm{co}}}\right) ^{3/2}=x_{0}^{3/2}\omega
_{\ast }\;.  \label{fastness}
\end{equation}
Here, $r_{0}$ is the radius where $\Omega $ reaches its maximum and $
x_{0}\equiv r_{0}/r_{\mathrm{in}}$. Using equation (\ref{fastness}), we
obtain $\omega _{\mathrm{s,c}}\cong 0.75$ for $\omega _{\ast \mathrm{,c}
}=0.5 $ from Figure~\ref{fig2}$a$ and $\omega _{\mathrm{s,c}}\cong 0.83$ for $\omega
_{\ast \mathrm{,c}}=0.8$ from Figure~\ref{fig2}$b$. Our analysis indicates broad
transition zones for $j=0$ rather than narrow ones if $\omega _{\mathrm{s,c}
}<0.8$. The disk is Keplerian at all radii only for $\omega _{\ast \mathrm{,c
}}=\omega _{\mathrm{s,c}}=0.875$ and $x_{0}=x_{1}=1$.

We have extended our calculations to several binary X-ray pulsars observed
as QPO sources. One of the simplest and most frequently used models for QPOs
from X-ray pulsars is the beat frequency model \citep{AS85}. According to the
beat frequency model (BFM), the inhomogeneities at the inner edge of the
Keplerian disk modulate the X-ray intensity and the QPO is observed as the
beat frequency between the local Keplerian frequency and the neutron star
spin frequency,
\begin{equation}
\nu _{\mathrm{QPO}}=\nu _{\mathrm{K}}(r_{0})-\nu _{\ast }\;.  \label{bfm}
\end{equation}
The corresponding fastness parameter can be written as
\begin{equation}
\omega _{\mathrm{s}}^{\mathrm{BFM}}=\frac{1}{1+P_{\ast }\,\nu _{\mathrm{QPO}}
}\;,  \label{fastbfm}
\end{equation}
where $P_{\ast }$ is the spin period of the observed X-ray pulsar. The QPOs
detected from X-ray pulsars can be used, within the beat frequency model, to
obtain constraints on the fastness parameter. For each source, we assumed $
M_{\ast }=1.4\;M_{\odot }$ and $R_{\ast 6}=1$. We calculated $j$ as a
function of $\omega _{\ast }$ from equation (\ref{j}) and $\omega _{\mathrm{s
}}^{\mathrm{BFM}}$ from equation (\ref{fastbfm}) using the observed periods $
P_{\ast }$, period derivatives $\dot{P}_{\ast }$, X-ray luminosities $L_{
\mathrm{X}}$, and QPO frequencies $\nu _{\mathrm{QPO}}$ of these X-ray
pulsars \citep{T94,Wang96}. We
determined the mass accretion rates through $\dot{M}=L_{\mathrm{X}}R_{\ast
}/GM_{\ast }$. The magnetic dipole moments $\mu _{\ast 30}$ have been
recently deduced from the cyclotron features \citep{Cob02}. Sources like 4U $
1626-67$, which are known to be close to their spin equilibrium, indicate
lower values for the critical fastness parameter. The reason for this could
be the deviation of angular velocity from the Keplerian profile near the
inner disk regions \citep{LW96}. Figure~\ref{fig3}$a$ shows the predicted
rotation curve of 4U $1626-67$ when the pulsar was spinning up just before
the torque reversal in 1990. After the torque reversal, we expect the
angular motion of the disk matter around the same pulsar as in Figure~\ref{fig3}$b$.
The estimated angular velocity profiles of 4U $0115+63$ and Cen X-3 are
plotted in Figures~\ref{fig4} and 5, respectively. We could obtain the relevant
rotation profiles shown in Figures~\ref{fig3}, 4, and 5 only for specific values of $
\omega _{\ast }$ provided we have $\omega _{\mathrm{s}}=\omega _{\mathrm{s}
}^{\mathrm{BFM}}$ (see eq. [\ref{fastness}]). With these constraints, the disk
model depends on three parameters, $r_{\mathrm{in}}$, $\beta $ and $j$,
and there is a unique solution, such that if one of these three parameters
is known, the other two can be obtained as the parameters that yield
the unique solution for $\Omega (r)$. For example, if the value of $r_{\mathrm{in}}$
is assumed, e.g., $r_{\mathrm{in}}=r_{\mathrm{A}}$, then $\beta $ and
$j$, therefore the torque on the star can be obtained; numerically by
scanning the $\beta, j$ parameter space until the unique solution is found.
In the applications, we used observed values of $\dot{\Omega}$ to
determine $j$ and scanned in $\beta $ only, to use a shortcut to the model
solution. We now summarize our results for each of these individual sources.

4U $1626-67$ is a low-mass X-ray binary pulsar with a $7.66$ s spin period.
The pulsar was in a steady spin-up state with $\dot{P}_{\ast }\cong
-4.86\times 10^{-11}$ s s$^{-1}$ during $1977-1990$. After the 1990 torque
reversal, the source underwent steady spin-down with $\dot{P}_{\ast }\cong
4.32\times 10^{-11}$ s s$^{-1}$. Although there is no indication for an
abrupt change in the X-ray luminosity at the torque reversal, a gradual
decrease over a decade in the mass accretion rate can be seen from the
archival flux history \citep{Chak97}. The QPO frequency is $\sim 0.04$
Hz during both the spin-up and spin-down episodes. The estimated mass
accretion rates are $\dot{M}_{17}\approx 1$ for spin-up and $\dot{M}
_{17}\approx 0.5$ for spin-down. The beat frequency model gives $\omega _{
\mathrm{s}}^{\mathrm{BFM}}=0.76$. Figure~\ref{fig3}$a$ shows the predicted rotation
curve, when the pulsar spins up, for $\omega _{\mathrm{s}}\cong 0.76$ with $
\omega _{\ast }=0.65$ and $\beta =6.48$. The angular velocity profile during
the spin-down episode was estimated for the same fastness parameter as in
Figure~\ref{fig3}$b$ with $\omega _{\ast }=0.2$ and $\beta =163$. We find $r_{\mathrm{
in}}/r_{\mathrm{co}}\cong 0.75$ and $r_{\mathrm{in}}/r_{0}\cong 0.91$ for
spin-up, and $r_{\mathrm{in}}/r_{\mathrm{co}}\cong 0.34$ and $r_{\mathrm{in}
}/r_{0}\cong 0.41$ for spin-down with $r_{\mathrm{co}}\cong 6.5\times 10^{8}$
cm. At the torque reversal, we expect that $\gamma _{\phi }\,s_{\mathrm{eff}
}^{2}$ changes from $4.7$ to $3.8$ according to our values of $\beta $ and $
\omega _{\ast }$ for $\mu _{\ast 30}=2.2$. Note that the angular velocity
transition zone is quite broad for the spin-down state, i.e., $\delta
r/r_{0}=1-r_{\mathrm{in}}/r_{0}\cong 0.59$ in comparison with $\delta
r/r_{0}\cong 0.09$ during the spin-up episode.

4U $0115+63$ is a $3.61$ s pulsar. It has $\dot{P}_{\ast }\cong -3.66\times
10^{-12}$ s s$^{-1}$ and $\dot{M}_{17}\approx 0.43$. The observed QPO
frequency is $0.062$ Hz. We inferred a fastness parameter of $0.817$ from
the beat frequency model. We find $r_{0}\cong 3.4\times 10^{8}$ cm and $r_{
\mathrm{co}}\cong 3.9\times 10^{8}$ cm. The same value for $\omega _{\mathrm{
s}}$ can be determined from the angular velocity profile if $\omega _{\ast
}=0.78$ (see Fig.~\ref{fig4}). We get $r_{\mathrm{in}}/r_{0}\cong 0.97$. The width of
the transition zone can be estimated as $\delta r/r_{0}\cong 0.03$. An Alfv\'{e}n
radius of $4.3\times 10^{8}$ cm for $\mu _{\ast 30}\cong 1$ gives $r_{
\mathrm{in}}/r_{\mathrm{A}}\cong 0.77$. The solution in Figure~\ref{fig4} was
obtained for $\beta =4.25$. This further implies that $\gamma _{\phi }\,s_{
\mathrm{eff}}^{2}\cong 1.7$.

Cen X-3 is a $4.83$ s pulsar. It has $\dot{P}_{\ast }\cong -4.29\times
10^{-11}$ s s$^{-1}$ and $\dot{M}_{17}\approx 4.3$. The observed QPO
frequency is $0.035$ Hz. The fastness parameter inferred from equation (\ref
{fastbfm}) is $0.855$. We estimate $r_{0}/r_{\mathrm{co}}\cong 0.9$. A
boundary layer width of $\delta r/r_{0}\cong 0.01$ can be deduced from $r_{
\mathrm{in}}/r_{0}\cong 0.99$. Our computation gives $\omega _{\mathrm{s}
}=0.855$ only for $\omega _{\ast }=0.84$ and $\beta =3.92$ (see Fig.~\ref{fig5}).
This yields $r_{\mathrm{in}}\cong 4.27\times 10^{8}$ cm and $\gamma _{\phi
}\,s_{\mathrm{eff}}^{2}\cong 14$ for $\mu _{\ast 30}\cong 2$ and $r_{\mathrm{
co}}\cong 4.8\times 10^{8}$ cm. Our model predicts $r_{\mathrm{in}}/r_{
\mathrm{A}}\cong 1.28$ for this source.

\section{MODEL APPLICATIONS IN COMPARISON WITH RESULTS OF NUMERICAL SIMULATIONS}\label{sec5}

The recent numerical study made by RUKL02 is the only work where the
slow viscous accretion and the disk-magnetized star interaction were
investigated in full two-dimensional time-dependent simulations with
quasi-stationary conditions. The results of RUKL02 are relevant to all
stationary models that describe the magnetic interaction of an accreting star
with its surrounding disk. As the simulations by RUKL02 are
two-dimensional whereas our work is limited to one-dimension, we
expect significant differences between some of our results and
their predictions. The inner boundary chosen by RUKL02 to investigate the
dynamics of disk accretion is the rotating star on which free boundary
conditions are applied to several hydrodynamic variables. In our work,
the computational region is limited by the inner radius of the disk where
the disk plasma corotates with the stellar magnetosphere. We keep the
Keplerian law fixed as the asymptotic outer boundary condition
on the rotation rate of the disk matter. The outer boundary
conditions taken by RUKL02 on all hydrodynamic variables are fixed
for the maximum computational region and free for a smaller simulation region (inner disk).

Although our boundary conditions differ significantly from the ones used by RUKL02,
it is possible to make appropriate conversions between our reference values and those of
RUKL02 for the inner disk radius and the physical quantities such as the stellar rotation rate
and the angular momentum flux. Our reference value for the distance is always the true inner radius
of the disk, $r_{\mathrm{in}}$, which is equivalent to one in dimensionless units.
The innermost disk radius within which all the disk matter goes up into the funnel flow
changes in time relative to the reference value for the distance, $R_{0}$, in numerical
simulations of RUKL02. This innermost radius corresponds to $r_{\mathrm{in}}$ in our model
since the matter nearly corotates with the star inside $r_{\mathrm{in}}<R_{0}$.

The dimensionless stellar rotation rate, that is the fastness parameter in our units is given by
\begin{equation}
\omega _{\ast }\equiv \omega _{\ast 0}\left[ \frac{\Omega _{\mathrm{K}
}(R_{0})}{\Omega _{\mathrm{K}}(r_{\mathrm{in}})}\right] =\omega _{\ast
0}\left( \frac{r_{\mathrm{in}}}{R_{0}}\right) ^{3/2}\;,  \label{omgcomp}
\end{equation}
where $\omega _{\ast 0}$ is the fastness parameter in units of RUKL02.

The dimensionless total angular momentum flux to the star in our units can be written as
\begin{equation}
j\equiv \frac{\dot{J}}{\dot{M}r_{\mathrm{in}}^{2}\Omega _{\mathrm{K}}(r_{
\mathrm{in}})}=\varphi \left( \frac{\dot{M}_{0}}{\dot{M}}\right) \left( 
\frac{R_{0}}{r_{\mathrm{in}}}\right) ^{1/2}\;,  \label{fluxcomp}
\end{equation}
where $\dot{J}=\varphi \dot{M}_{0}R_{0}^{2}\,\Omega _{\mathrm{K}}(R_{0})$ is the total
angular momentum flux to the star and $\dot{M}_{0}$ is the reference mass accretion rate
in units of RUKL02.

In the following, we will estimate the approximate location of the braking radius $r_{\mathrm{br}}$,
using our model parameters $\beta$, $j$, and $\omega _{\ast}$. The braking radius was first
defined by RUKL02 as the radius inside which the disk is significantly disturbed
by the stellar field. In almost all simulations by RUKL02 (especially the relevant ones with
``Type I'' initial conditions), the density in the disk drops sharply at $r\lesssim r_{\mathrm{br}}$
and the matter is magnetically braked for $r_{\mathrm{in}}<r\lesssim r_{\mathrm{br}}$.
The density in the disk rises again near the innermost radius. The radial variation of density at
each time step of these numerical simulations is qualitatively reminiscent of the behavior of our
dynamical viscosity, $f(x)=3\pi \nu \Sigma /\dot{M}$ (see Fig.~\ref{fig1}). To make a
quantitative comparison of our model predictions with the results of RUKL02,
we use the numerical data of simulations for accretion to a slowly rotating
star with $\omega _{\ast 0}=0.19$. We select the numerical values given by RUKL02 for the
mass accretion rate $\dot{M}/\dot{M}_{0}$, the angular momentum flux $\varphi$, and the inner disk radius
$r_{\mathrm{in}}/R_{0}$ at times $t=30P_{0}$ and $t=40P_{0}$, where $P_{0}$ is the
rotational period at $R_{0}$. Using equations (\ref{omgcomp}) and (\ref{fluxcomp}),
we calculate our model parameters $\omega _{\ast}$ and $j$. Equations (\ref{dvis}) and (\ref{expt})
can be employed together with the condition $f(x_{1})\ll 1$ (see \S \S\ \ref{sec3} and \ref{sec4})
to solve for $\beta$. Once all model parameters are determined, the braking radius
$r_{\mathrm{br}}\approx (r_{\mathrm{in}}/R_{0})x_{1}$ where the density becomes minimum
can be readily found from equation (\ref{expt}). Our numerical calculations yield
$r_{\mathrm{br}}\approx 2.61$ at $t=30P_{0}$ and $r_{\mathrm{br}}\approx 2.65$ at $t=40P_{0}$.
These estimates are in agreement with the results of RUKL02.

\section{DISCUSSION}\label{sec6}

\begin{figure}
\epsscale{1.0} \plottwo{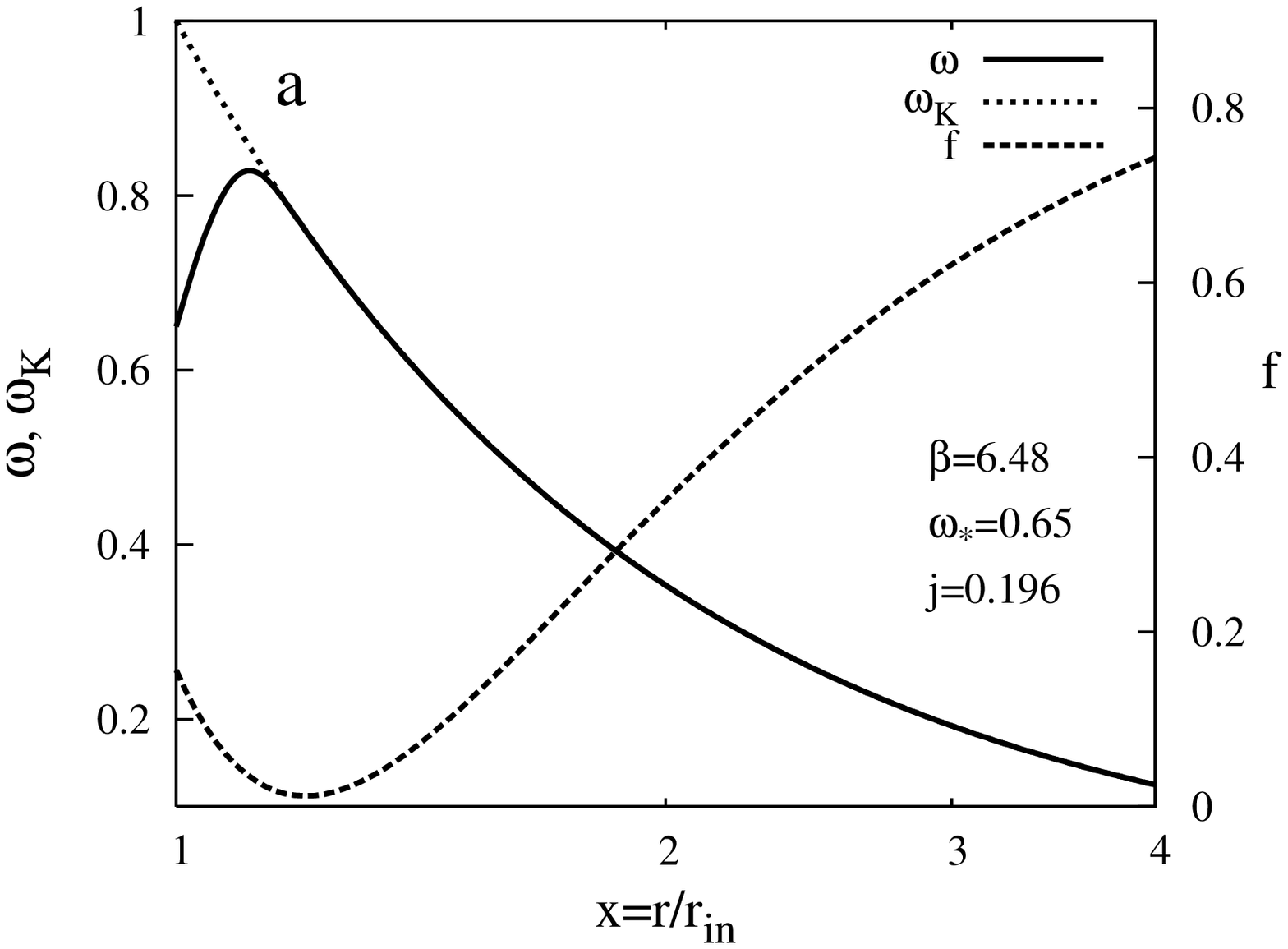}{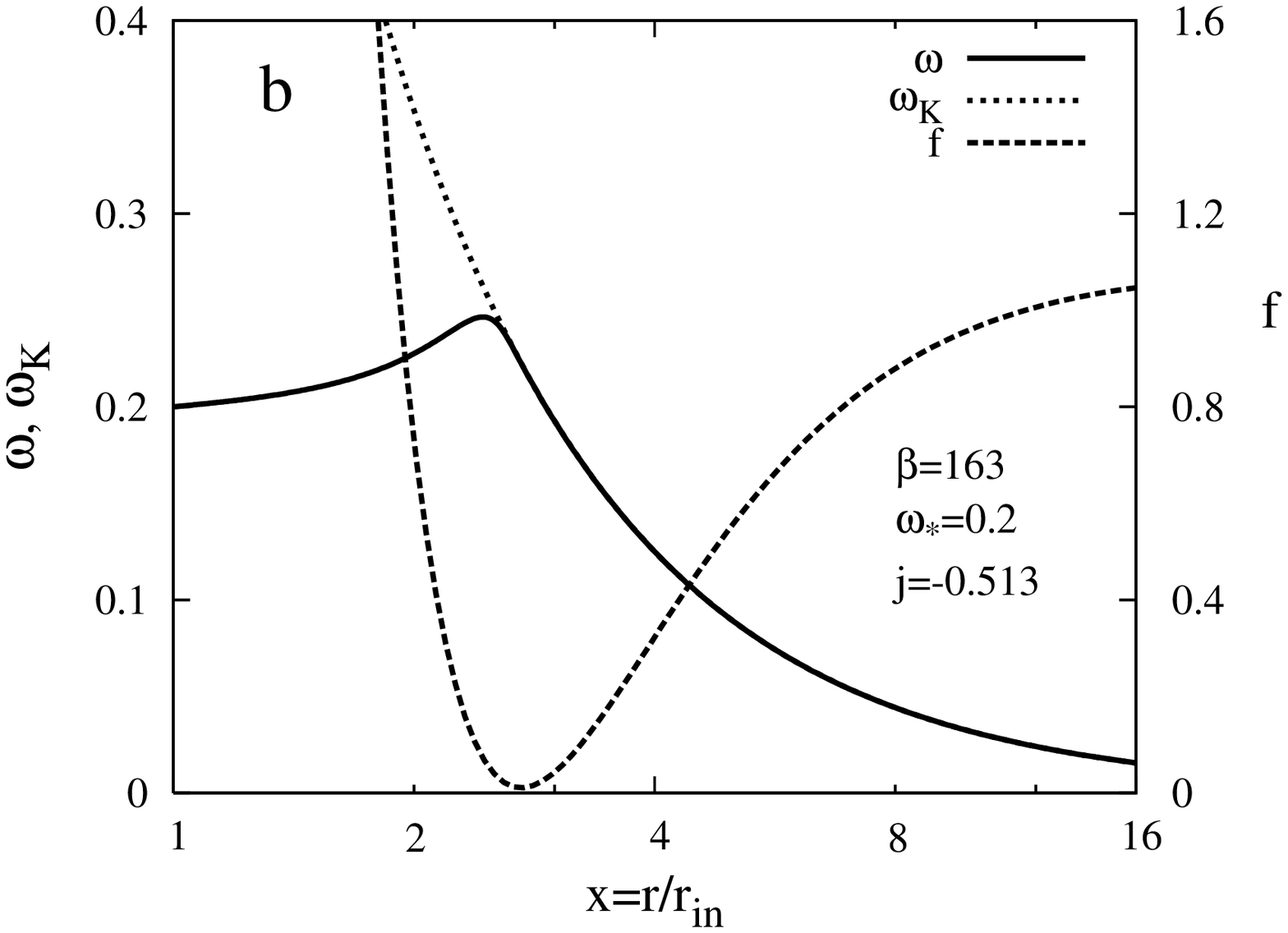} 
\caption{Estimated rotation rate profiles $\omega (x)\equiv \Omega /\Omega _{\mathrm{K}}(r_{\mathrm{in}})$ and
the corresponding normalized vertically integrated dynamical viscosities $f(x)=3\pi \nu \Sigma /\dot{M}$ for a
magnetically threaded disk around an X-ray pulsar with observed parameters appropriate for 4U $1626-67$
before and after the torque reversal. The panel (a) represents the angular velocity transition zone during the spin-up
episode. The transition region is broad as seen from panel (b) when the source is in spin-down state. The rotation
rate profiles $\omega (x)$ shown by \emph{solid curves} in panels (a) and (b) obtain for
$\beta =6.48$, $\omega_{\ast }=0.65$, $j\simeq0.196$ and $\beta =163$, $\omega_{\ast }=0.2$, $j\simeq-0.513$,
respectively. In each panel the \emph{dashed curve} corresponds to $f(x)$ and the \emph{dotted curve} to the Keplerian
profile $\omega _{\mathrm{K}}(x)=x^{-3/2}$, where $x\equiv r/r_{\mathrm{in}}$. \label{fig3}}
\end{figure}

\begin{figure}
\epsscale{1.0} \plotone{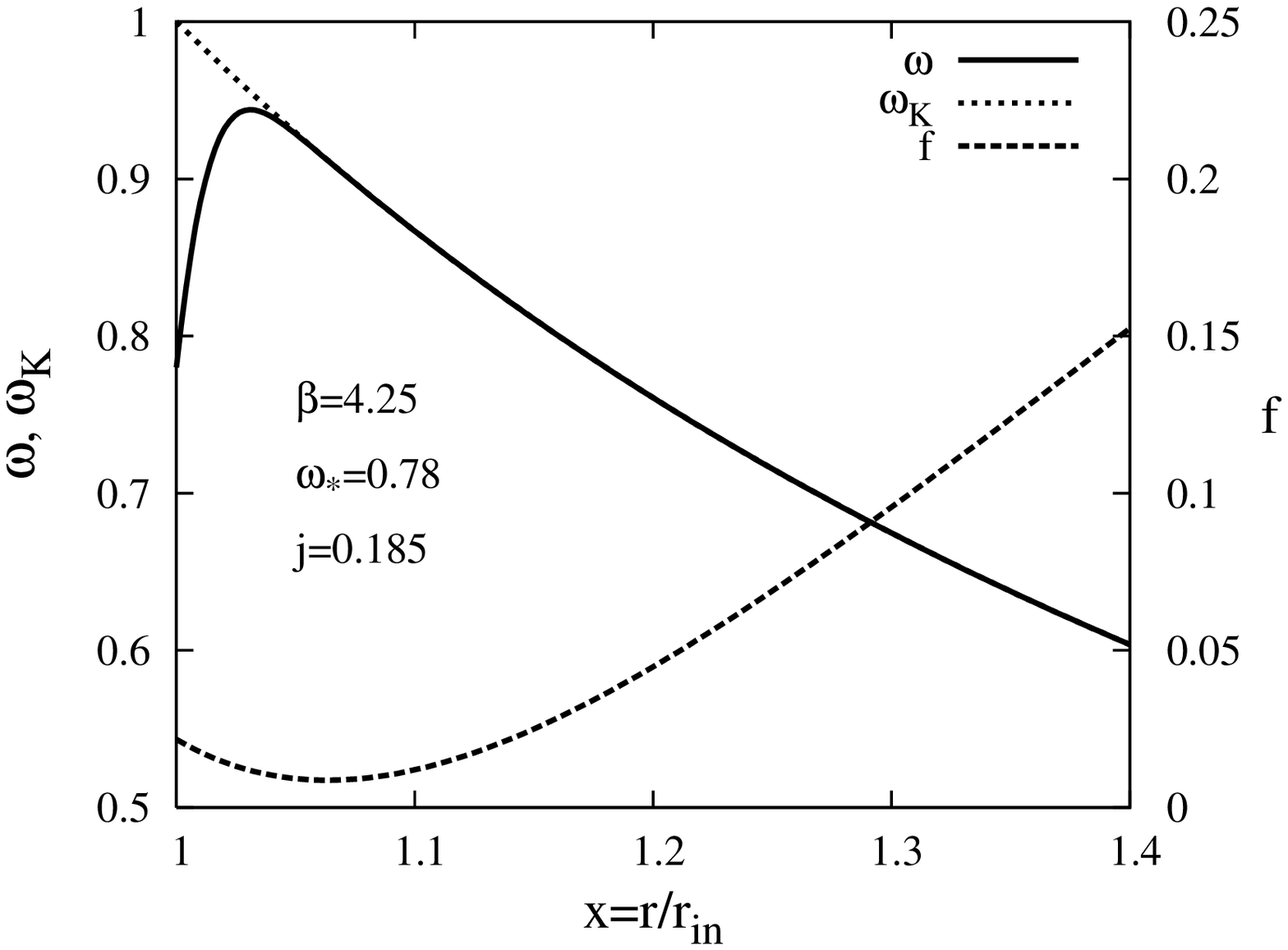} 
\caption{Radial variation of angular velocity $\omega (x)\equiv \Omega /\Omega _{\mathrm{K}}(r_{\mathrm{in}})$ and
the corresponding normalized vertically integrated dynamical viscosity $f(x)=3\pi \nu \Sigma /\dot{M}$ for a disk around
an X-ray pulsar with observed parameters appropriate for 4U $0115+63$. The rotation rate profile $\omega (x)$
shown by the \emph{solid curve} obtains for $\beta =4.25$, $\omega_{\ast }=0.78$, $j\simeq0.185$. The \emph{dashed curve}
corresponds to $f(x)$ and the \emph{dotted curve} to the Keplerian profile $\omega _{\mathrm{K}}(x)=x^{-3/2}$, where
$x\equiv r/r_{\mathrm{in}}$. \label{fig4}}
\end{figure}

\begin{figure}
\epsscale{1.0} \plotone{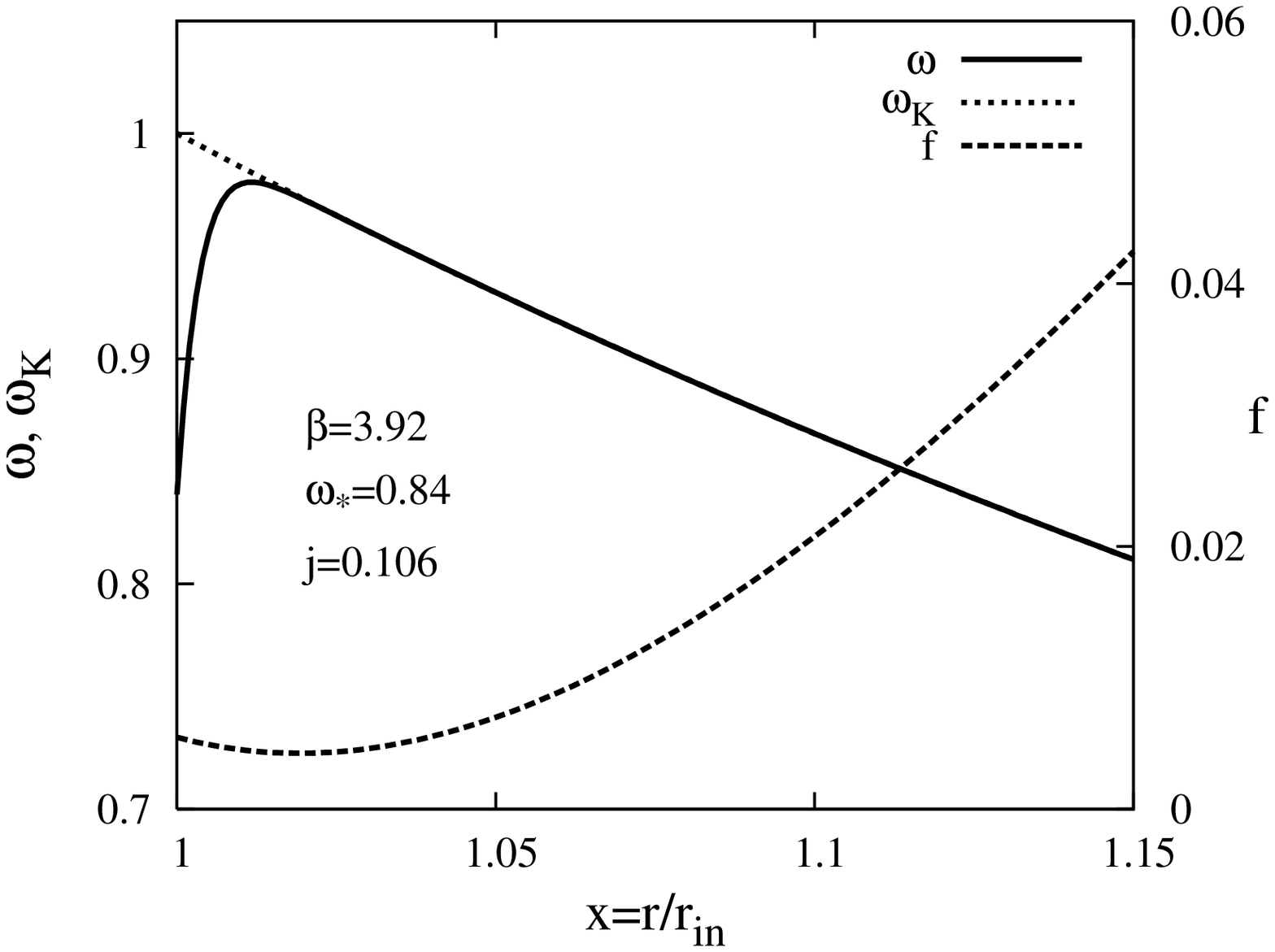} 
\caption{Radial variation of angular velocity $\omega (x)\equiv \Omega /\Omega _{\mathrm{K}}(r_{\mathrm{in}})$ and
the corresponding normalized vertically integrated dynamical viscosity $f(x)=3\pi \nu \Sigma /\dot{M}$ for a disk around
an X-ray pulsar with observed parameters appropriate for Cen X-3. The rotation rate profile $\omega (x)$
shown by the \emph{solid curve} obtains for $\beta =3.92$, $\omega_{\ast }=0.84$, $j\simeq0.106$. The \emph{dashed curve}
corresponds to $f(x)$ and the \emph{dotted curve} to the Keplerian profile $\omega _{\mathrm{K}}(x)=x^{-3/2}$, where
$x\equiv r/r_{\mathrm{in}}$. \label{fig5}}
\end{figure}

We investigated the rotational dynamics of an accretion disk threaded by the
dipolar magnetic field of a neutron star with constant screening factor.
Within model assumptions, we found that a range of narrow or wide angular
velocity transition zones with non-Keplerian rotation are consistent with
observations. Our approach is analogous to the work of Glatzel (1992) for
the non-magnetic case. We derived a vertically integrated dynamical
viscosity form for a magnetically threaded disk from the conservation of
mass and angular momentum using the Keplerian rotation as a particular
solution for the angular motion of the disk plasma. Assuming that the whole
disk can be treated by the same dynamical viscosity, we were able to obtain
rotation curves without the restricting assumption of a thin boundary layer.
We also used a more realistic, spatially varying azimuthal pitch, $B_{\phi
}^{+}/B_{z}$, throughout the angular velocity transition zone instead of
keeping it constant as in the model of Ghosh \& Lamb. The effect of the
magnetically enhanced viscous stress is included in our calculations. The
rotation of the disk matter is almost Keplerian outside the corotation
radius. The deviation from the Keplerian law is noticeable where the
effective viscous stress is negligible as compared with the magnetic stress.
This can be realized at a specific disk radius, $r_{1}=x_{1}r_{\mathrm{in}}$
, where the vertically integrated dynamical viscosity function, $f$, goes
through a local minimum and actually nearly vanishes. The angular velocity
at $r_{1}$ is still nearly Keplerian. We find that the angular motion of the
inner disk plasma from $r_{1}$ to $r_{\mathrm{in}}$ is controlled by both
the viscous and magnetic stresses. The true inner disk radius, $r_{\mathrm{in
}}$, is therefore determined in our model by the close balance between the
magnetic stress and the viscous and material stresses. The width of the
transition zone, $\delta \lesssim x_{1}-1$, depends basically on the model
parameters $\beta $, $j$, and $\omega _{\ast }$ (see \S \S\ \ref{sec3} and
\ref{sec4}). The physical meaning of this is that for different values of
$\dot{M}$, $\Omega _{\ast }$ and for given $\mu _{\ast }$, the torque on the
star from a disk threaded by a screened dipole field can be achieved with
the adjustment of the $\Omega (r)$ profile through a transition zone. The
rotation is nearly Keplerian throughout the whole disk and the angular
velocity transition region is narrow in the case of dynamical viscosity
decreasing towards the innermost edge of the disk for the current values of
$\beta $, $j$, and $\omega _{\ast }$. The boundary layer approximation of
GL79 remains valid only for $f(x\simeq 1)\approx 0$ (see Fig.~\ref{fig5}). In this
sense, the viscously unstable inner disk solutions of BC98 and CH98 could be
an artefact of taking a fully Keplerian disk. In general the inner disk is not
Keplerian; the angular momentum transfer with a dynamical viscosity increasing
at small radii cannot satisfy the boundary condition imposed by the rotating
magnetosphere if the inner disk is assumed to be Keplerian (see Fig.~\ref{fig3}$b$).

Using observed parameters of several sources in equilibrium, spin-up or
spin-down states, we found that the non-Keplerian transition region for a
Keplerian flow to match to the rotation rate of the stellar magnetosphere
can be broad or narrow depending on source state. The beat frequency
interpretation of QPOs provides a constraint on the fastness parameter for a
given source, and thus on the model parameters ($\beta $, $j$, $\omega
_{\ast }$) that determine the appropriate rotation curve and the width of
the transition zone. Among the examples we considered, the transition zones
are narrowest, $\delta r/r_{0}\sim 0.01-0.1$ for sources in spin-up (\objectname{4U 
$1626-67$}, \objectname{4U $0115+63$}, \objectname{Cen X-3}); 
while $\delta r/r_{0}\sim 0.02-0.24$ for sources in rotational equilibrium; and
broadest, $\delta r/r_{0}\sim 0.6$ for \objectname{4U $1626-67$}, when the source
is in spin-down state. QPOs from disk-fed torque reversing X-ray binary pulsars may
elucidate the function of non-Keplerian accretion flows and/or the width of the transition
zones in the spin behavior of these sources.

Results of the numerical simulations by RUKL02 were compared to our model.
These two-dimensional simulations describe an analogous situation where the disk is
threaded by the magnetic field of the star; both viscosity and diffusivity are of the same
order of magnitude; and the disk matter slowly accretes across the field lines. Although
the direct comparison of the angular velocity distributions is not possible because of
different conditions assumed at the boundaries of computational domains; there are
remarkable similarities between our model predictions and the results of RUKL02
(see \S\ \ref{sec5}).

\acknowledgments
We thank the anonymous referee for constructive comments and suggestions,
K. Y. Ek\c{s}i, F. K. Lamb, and H. Spruit for useful discussions,
and E. N. Ercan for her support and encouragement. This work was supported
by Bo\u{g}azi\c{c}i University Research Foundation under code 01B303D for M. H. E.,
by  the Sabanc\i\ University Astrophysics and Space Forum,
by the High Energy Astrophysics Working Group of T\"{U}B\.{I}TAK
(The Scientific and Technical Research Council of Turkey)
and by the Turkish Academy of Sciences for M. A. A.

\clearpage

\clearpage

\clearpage

\clearpage

\clearpage

\end{document}